\documentclass{ws-procs9x6}

\newcommand{\refeq}[1]{(\ref{#1})}
\def\etal {{\it et al.}}

\begin{document}

\title{ANTIHYDROGEN, CPT, AND NATURALNESS%
\footnote{Based on an invited talk at CPT'13 --  the Sixth Meeting on CPT and Lorentz Symmetry, Bloomington, Indiana, June 17-21, 2013.}
}

\author{MAKOTO C.\ FUJIWARA}

\address{TRIUMF National Laboratory for Particle and Nuclear Physics\\
Vancouver, British Columbia, V6T 2A3, Canada\\
and\\
Department of Physics and Astronomy,
University of Calgary\\
Calgary, Alberta, T2N 1N4, Canada\\
E-mail: Makoto.Fujiwara@triumf.ca}


\begin{abstract}
Studying fundamental symmetries of Nature has proven fruitful in particle physics. I argue that recent results at the LHC, and the naturalness problem highlighted by them, provide a renewed motivation for tests of CPT symmetry as a probe for physics beyond quantum field theory. I also discuss prospects for antihydrogen CPT tests with sensitivities to Planck scale suppressed effects.

\end{abstract}

\bodymatter

\section{Introduction}

It is clear that testing CPT invariance at highest possible precision is a worthwhile effort, given its fundamental importance in modern physics.\cite{SME} Comparisons of antihydrogen atoms ($\rm {\overline{H}}$) with their well-studied matter counterpart, atomic hydrogen ($\rm {H}$), could provide competitive tests of CPT. (See, e.g., Ref.\ \refcite{AIP}). In this short paper, I will try to put $\rm {\overline{H}}$ studies in the larger context of current particle physics, and argue that the recent LHC results provide enhanced motivations for symmetry tests with $\rm {\overline{H}}$.

\section{Naturalness and CPT}
Before going into antihydrogen, let us digress and ask a rather basic question: what is particle physics? According to Grossman\cite{Grossman}, particle physics is about asking a simple question:
\begin{equation}
\mathcal{L} = ?
\label{eq1}
\end{equation}
To this simple question, we seem to have a simple answer: the Standard Model (SM), including the recently discovered final piece, a Higgs boson. However, as is well known, the SM has a number of open issues, not least of which is the (technical) naturalness problem. Loosely speaking, the naturalness (also called the hierarchy, or the fine-tuning) problem in this context refers to the lightness of the observed Higgs mass compared to the Planck (or the grand unification) scale, which implies fine tuning of the Higgs parameters by some $O(30)$ within the SM.

In the past decades, the issue of naturalness has been a central guiding principle in particle physics. A number of solutions to the problem have been proposed and studied extensively, the most popular scenario being supersymmetry. These solutions usually require that new phenomena appear at the energy scale near the electroweak scale, leading to the expectation that we would observe new physics beyond the SM at the LHC. The lack of such observations thus far has ruled out many of the most attractive new physics scenarios, and it seems to be putting many particle physicists in the state of `soul searching.' Of course, there is still room for discoveries as the LHC energy is increased, which may solve the naturalness problem, but most of the surviving scenarios appear rather contrived. In addition to the light Higgs mass, the cosmological constant presents an even greater challenge to our belief in naturalness, naively requiring a fine-tuned cancelation at the $O(120)$ level.

This apparently disparate situation has lead to the increasing popularity of the anthropic principle, which states certain parameters in physics are fine tuned (possibly in a landscape of the `multiverse') to allow the existence of the observer.
%
Before accepting this controversial (but logically possible) option, however, I wish to step back and ask: are we asking the right question? When the answer we get (i.e., incredible degrees of fine tuning) does not make sense, it is possible that we are asking a wrong question.

In fact, implicit in the question in Eq.~\refeq{eq1} is the validity of quantum field theory (QFT), at least as the low energy effective description of Nature. However, it is the effective QFT framework itself that gives rise to the fine-tuning problems in the first place. The question we should be asking may be about the validity of the framework itself, rather than the ingredients in it. The possibility of such an option, although admittedly speculative, motivates putting the QFT framework to stringent experimental tests.

How does one test the validity of QFT? One possibility is to improve the precision of measurements of physical quantities that can be predicted precisely. The electron $g-2$ factor is a leading candidate. However, as a test of QED, it is currently limited by independent knowledge of the fine structure constant. Another approach is to search for a violation of symmetries guaranteed in QFT, such as CPT. $\rm {\overline{H}}$-$\rm {H}$ comparisons fall into the latter category. By directly comparing matter and antimatter, we would be free from the uncertainties due to theory or the constants.

Search for violation of CPT and/or Lorentz invariance has been the subject of considerable recent activities. A general effective field theory framework by Kosteleck\'{y} \etal, known as the Standard-Model Extension (SME), has provided an extremely powerful framework in which to test these symmetries.\cite{SME} It has been well known that the CPT theorem is guaranteed in QFT under general assumptions including Lorentz invariance, locality, and unitarity. A notable finding in the past decade is that within the SME (and presumably in any local QFT), CPT violation is always accompanied by Lorentz violation. This implies that limits on Lorentz violation in matter-only experiments also provide constraints on CPT violation. Given that extremely sensitive limits of the SME parameters are being placed via matter-only experiments\cite{SME}, is there need for antimatter experiments?

Matter-antimatter comparison experiments confront the entire framework of (effective) QFT by testing CPT-odd, but Lorentz-even interactions, which are forbidden in local QFT. In other words, any violation of CPT, e.g., in comparison of $\rm {\overline{H}}$ and $\rm {H}$ atomic spectra, would indicate physics beyond QFT, forcing a fundamental change in our understanding of Nature.

Of course, it is by no means guaranteed that any potential modifications of QFT would lead to an observable CPT violating effect in  $\rm {\overline{H}}$ experiments. However, antihydrogen should serve at least as a `lamp post' case, because of the potentially very high sensitivities it could offer.\cite{AIP}

Since the CPT'13 meeting, I learned that there are in fact attempts to explain the fine-tuning problem using (mildly) nonlocal theory beyond the conventional QFT, e.g., by considering effects of wormholes in the multiverse.\cite{non-local} Whether this class of models would have observable implications at low energies is an open question.

\section{Antihydrogen experiments at the CERN AD}

Several experiments are ongoing or under construction at the CERN Antiproton Decelerator, with the goal of testing fundamental symmetries with $\rm {\overline{H}}$. Here, I briefly discuss one of the experiments, ALPHA (Antihydrogen Laser PHysics Apparatus). Since the CPT'10 meeting, ALPHA has made significant progress. Indications of trapped $\rm {\overline{H}}$, which I reported at CPT'10,\cite{mcf} has now been unambiguously confirmed.\cite{trapping} The confinement times have been extended to as long as 1000 seconds\cite{longtime}. First proof-of-principle demonstration of spectroscopic measurement on $\rm {\overline{H}}$ has been performed by driving hyperfine transitions via microwaves\cite{microwaves}. An entirely new trap (ALPHA-2) has been constructed and is being commissioned to allow laser and improved microwave spectroscopy.

Laser spectroscopy of the 1s-2s level is the golden mode for a CPT test with $\rm {\overline{H}}$, since the same transition in $\rm {H}$ is measured to parts in 10$^{-15}$. Hyperfine spectroscopy offers a complementary test.\cite{Widmann,AIP} In this respect, recent developments in measurements of the bare antiproton $g$-factor are encouraging. While the latter probes the long-distance magnetic property of the antiproton, $\rm {\overline{H}}$ hyperfine splitting can probe the antiproton's internal structure via the contact interaction of the positron and the antiproton.\cite{Widmann} Recall that while the proton electric charge is very well known, there is a puzzle in its charge distribution. The gravitational interaction of antimatter is another increasingly active subject. See, e.g., Ref.\ \refcite{gravity}.

A benchmark for CPT tests may be the sensitivity to Planck scale suppressed effects, $\Delta E \sim m_{\rm proton}^2/M_{\rm Planck} \sim 10^{-18}$ GeV.\cite{AIP} In frequency units, this corresponds to the precision of $\Delta f \sim 100$ kHz. This is within the reach of current antihydrogen experiments.


\section*{Acknowledgments}
I wish to thank Alan Kosteleck\'{y} and the organizing committee for a stimulating meeting. I thank the members of ALPHA for fruitful collaboration, Art Olin and Dave Gill for a critical reading of the manuscript. This work is supported in part by Canada's NSERC, and TRIUMF.

\end{document}